\documentclass[preprint,showpacs,preprintnumbers,amsmath,amssymb]{revtex4}
\usepackage{booktabs}
\usepackage{mathrsfs}
\usepackage{epsfig}
\usepackage{graphicx}
\usepackage{dcolumn}
\usepackage{bm}
\usepackage{amsmath}
\usepackage{slashed}       
\usepackage{amssymb}
\usepackage{amsfonts}
\usepackage{graphics,float,psfrag}   
\usepackage[usenames,dvips]{color}
\usepackage{rotating}

\let\jnfont=\rm
\def\NPB#1,{{\jnfont Nucl.\ Phys.\ B }{\bf #1},}
\def\PLB#1,{{\jnfont Phys.\ Lett.\ B }{\bf #1},}
\def\EPJC#1,{{\jnfont Eur.\ Phys.\ Jour.\ C }{\bf #1},}
\def\PRD#1,{{\jnfont Phys.\ Rev.\ D }{\bf #1},}
\def\PRL#1,{{\jnfont Phys.\ Rev.\ Lett.\ }{\bf #1},}
\def\MPLA#1,{{\jnfont Mod.\ Phys.\ Lett.\ A }{\bf #1},}
\def\JPG#1,{{\jnfont J.\ Phys.\ G}{\bf #1},}
\def\CTP#1,{{\jnfont Commun.\ Theor.\ Phys.\ }{\bf #1},}
\def\ZPC#1,{{\jnfont Z.\ Phys.\ C }{\bf #1},}
\def\JHEP#1,{{\jnfont JHEP \ }{\bf #1},}
\def\lsim{\raise0.3ex\hbox{$<$\kern-0.75em\raise-1.1ex\hbox{$\sim$}}}
\def\gsim{\raise0.3ex\hbox{$>$\kern-0.75em\raise-1.1ex\hbox{$\sim$}}}

\begin{document}
\preprint{\parbox{1.2in}{\noindent arXiv:}}

\title{SUSY induced top quark FCNC decay $t \to c h$ after Run I of LHC}

\author{Junjie Cao$^{1,2}$, Chengcheng Han$^3$, Lei Wu$^4$, Jin Min Yang$^3$, Mengchao Zhang$^3$}

\affiliation{
  $^1$  Department of Physics,
        Henan Normal University, Xinxiang 453007, China \\
  $^2$ Center for High Energy Physics, Peking University,
       Beijing 100871, China \\
  $^3$ State Key Laboratory of Theoretical Physics,
      Institute of Theoretical Physics, Academia Sinica, Beijing 100190, China \\
  $^4$ ARC Center of Excellence for Particle Physics at the Terascale, \\
      School of Physics, The University of Sydney, NSW 2006, Australia
      \vspace{1cm}}

\begin{abstract}
In light of the Higgs discovery and the nonobservation of sparticles at the LHC,
we revisit the SUSY induced top quark flavor changing decay into the Higgs boson.
We perform a scan over the relevant SUSY parameter space by considering the constraints 
from the Higgs mass measurement, the LHC search for SUSY,
the vacuum stability, the precision electro-weak observables as well as
$B \to X_s \gamma$.  We have the following observations:
(1) In the MSSM, the branching ratio of $ t \to c h$ can only reach
$3.0\times10^{-6}$, which is about one order smaller than previous results obtained before
the advent of the LHC. Among the considered constraints, the Higgs mass and
the LHC search for sparticles are found to play an important role in limiting the prediction.
(2) In the singlet extension of the MSSM, since the squark sector
is less constrained by the Higgs mass, the branching ratio of $t \to c h$ can reach the order of $10^{-5}$ in the allowed parameter space.
(3) The chiral-conserving mixings $\delta_{LL}$ and $\delta_{RR}$ may have remanent effects on $t \to c h$ in heavy SUSY limit.
In the MSSM with squarks and gluino above 3 TeV and meanwhile the CP-odd Higgs boson mass around 1 TeV, the branching ratio of $t\to c h$
can still reach the order of $10^{-8}$ under the constraints.
\end{abstract}
\maketitle

\section{Introduction}

A scalar with mass around 125 GeV has been discovered at the LHC \cite{1207-a-com,1207-c-com}.
According to the analysis of the ATLAS and CMS collaborations, the measured properties of this scalar,
albeit with large experimental uncertainties, agree well with those of the Higgs boson in
the Standard Model (SM), which means that it plays a role in the electroweak (EW) symmetry breaking and also in the mass generation
for the fermions in the SM \cite{1303-atlas-Moriond,1303-c-com,1303-a-com}. Even so,
due to the deficiencies of the SM itself in describing the symmetry breaking,
it is well motivated to interpret this scalar in various frameworks of new physics.
Obviously, in order to ambiguously decipher the nature of the scalar, it is mandatory to
scrutinize both  experimentally and theoretically the couplings of the scalar,
including its self-interactions. In this direction, the
couplings of the scalar with the yet known heaviest particle, top quark, are of fundamental importance
since, as suggested by the LHC Higgs data, the $h\bar{t}t$ coupling is strong, and meanwhile it is widely
conjectured to be sensitive to new physics. In fact, great efforts have been paid recently to investigate the top-Higgs
associated production processes like $pp \to t\bar{t}h$ \cite{tth-theory,tth-LHC} and $p p \to q t h$ \cite{qth-theory}
at the LHC to extract the size and sign of the $h\bar{t}t$ Yukawa coupling, and also the top quark flavor changing decay
$t \to c h$ to prob anomalous top-Higgs interaction\cite{tch-theory,tch-LHC}.

Among the new physics models, the supersymmetric theory (SUSY) is a promising one due to its capability to solve the hierarchy
problem of the SM, unify the gauge coupling as well as provide a viable Dark Matter candidate\cite{MSSM,NMSSM}.
In SUSY, a SM-like Higgs boson $h$ around $125 {\rm GeV}$ usually implies third generation squarks at or heavier than
$1 {\rm TeV}$, and the preference of the heavy squarks is further corroborated by the
absence of any signal in the search for SUSY at the LHC. If the SUSY scale is really high, which was
focused on in many recent theoretical works\cite{Heavy-SUSY}, the only way to detect SUSY is through its possibly large remanent effects in EW processes.
Such effects may exist in the Higgs process because the dominant part of the Higgs couplings to squarks is proportional to soft SUSY breaking parameters\cite{MSSM}, and consequently, the suppression induced by the squark propagators in SUSY radiative correction to the process
may be compensated under certain conditions. This feature has been demonstrated in the SUSY correction to the $h\bar{b}b$
vertex\cite{SUSY-QCD-hbb}, the Higgs pair
production process at the LHC\cite{Cao-hh}, and also the Higgs rare decay $h \to \tau \bar{\mu}$\cite{SUSY-htaumu}. Here
we emphasize that the existence of the remanent effect in the asymptotic large SUSY mass limit does not contradict
with the Appelquist-Carazzone theorem\cite{Non-Decoup-theorem}, which is
valid only for supersymmetric theories with an exact gauge symmetry.
Previous studies on such remanent effects in SUSY, also see \cite{old-nondecoup}.

In this work, we focus on the top quark flavor changing decay $t \to c h$ in SUSY. The reasons that we are interested in it mainly
come from three considerations. Firstly, the LHC as a top factory has great capability to scrutinize the properties of top quark,
including its rare decay modes. As far as the flavor changing decay $t \to c h$ is concerned, its branching ratio in
the SM is only at the order of $10^{-14}$\cite{tch-SM}, while in SUSY it may be greatly enhanced to $10^{-4}$ according to previous studies\cite{cao-flavor-mixing,cao-top-fcnc}. Since any observation of the decay in future will be a robust evidence of new physics,
this decay should be paid attention to in the LHC era, especially noting the
fact that the Higgs boson has been recently discovered. Secondly, as introduced before, the LHC experiment has measured Higgs mass and pushed SUSY to
a rather high scale. These results have great impacts on the SUSY prediction about $t\to c h$, so it is necessary to update
previous studies on $t \to c h$ in light of the experimental progress. Thirdly,  unlike the other top FCNC processes in SUSY\cite{old-top-fcnc},
the decay $t \to c h$ may have remanent effect in heavy SUSY limit. From theoretical point of view,
it is worthwhile to investigate such a feature in detail. Besides, we remind that, if the flavor mixings between scharm and
stop are present, which may push up the rate of $t \to c h$ greatly, the LHC constraint on stop masses can be relaxed. This in return can
alleviate the fine tuning problem of the SUSY\cite{Flavored-natural}.

This paper is organized as follows. In section II, we parameterize the flavor mixings in squark sector and define our
conventions. We also list various constraints on SUSY. In section III, we study the decay $t \to c h$
in both low energy SUSY and heavy SUSY, and present some benchmark points at which the predictions on $t \to c h$ are optimized.
We also exhibit the features of the remanent effect on $t \to c h$.  Finally, we present
our conclusions in section IV.

\section{FCNC Interactions in SUSY}

In the supersymmetric theories such as the Minimal Supersymmetric Standard Model (MSSM)\cite{MSSM} and the Next-to Minimal Supersymmetric
Standard Model (NMSSM)\cite{NMSSM}, the squark sector consists of six up-type squarks
($\tilde u_L$, $\tilde c_L$, $\tilde t_L$, $\tilde u_R$, $\tilde
c_R$, $\tilde t_R$) and six down-type squarks ($\tilde d_L$, $\tilde s_L$, $\tilde b_L$,
$\tilde d_R$, $\tilde s_R$, $\tilde b_R$). In general,  the states with different chiral
and flavor quantum numbers in each type of squarks will mix to form mass eigenstates, and consequently
potentially large flavor changing interactions arise from the misalignment between
the rotations that diagonalize quark and squark sectors.
In the super-CKM basis, the $6\times 6$ squark mass matrix
${\cal M}^2_{\tilde q}$ ($\tilde q=\tilde u, \tilde d$) takes the
form \cite{susyf}
\begin{eqnarray}
{\cal M}^2_{\tilde q}=\left( \begin{array}{ll}
  (M^2_{\tilde q})_{LL}+ C_{\tilde q}^{LL} & (M^2_{\tilde q})_{LR}-C_{\tilde q}^{LR} \\
  \left( (M^2_{\tilde{q}})_{LR}-C_{\tilde q}^{LR}\right)^\dag
             & (M^2_{\tilde{q}})_{RR}+C_{\tilde q}^{RR} \end{array} \right) ,\label{sqmass}
\end{eqnarray}
where $C_{\tilde q}^{LL} = m_q^2 + \cos 2\beta M_Z^2 ( T_3^q - Q_q
s_W^2) \hat{\mbox{\large 1}}$, $C_{\tilde q}^{RR} = m_q^2 + \cos
2\beta M_Z^2 Q_q s_W^2 \hat{\mbox{\large 1}}$ and $C_{\tilde
q}^{LR}=m_q \mu (\tan \beta)^{- 2 T_3^q}$ are $3\times 3$ diagonal
matrices with $\hat{\mbox{\large 1}}$ standing for the unit matrix in
flavor space, $m_q$ being the diagonal quark mass matrix and
$T_3^q=\frac{1}{2},-\frac{1}{2}$ for $q=u,d$ respectively, and $\tan \beta =\frac{v_2}{v_1}$ is the ratio of the vacuum expectation
values of the $SU(2)$ doublet Higgs fields. If one only considers the flavor mixings between
the second and the third generation squarks, the soft breaking squared masses
$(M^2_{\tilde q})_{LL}$,  $(M^2_{\tilde q})_{LR}$ and
$(M^2_{\tilde q})_{RR}$ can be parameterized as
\begin{eqnarray}
(M^2_{\tilde{u}})_{LL} &= & \left ( \begin{array}{ccc}
    M_{Q_1}^2 &  0                           &  0 \\
    0         &  M_{Q_2}^2                   & \delta_{LL} M_{Q_2} M_{Q_3} \\
    0         &  \delta_{LL} M_{Q_2} M_{Q_3} & M_{Q_3}^2 \end{array}  \right ), \nonumber \\
 (M^2_{\tilde{u}})_{LR} &=&  \left (
\begin{array}{ccc}
0     &  0   &  0 \\
0     &  0   & \delta_{LR} v_2 M^{U}_{LR}\\
0     &  \delta_{RL} v_2 M^{U}_{RL}  & m_t A_t \end{array}  \right ), \nonumber \\
 (M^2_{\tilde{u}})_{RR} &= & (M^2_{\tilde{u}})_{LL}|_{M_{Q_i}^2 \to M_{U_i}^2,~ \delta_{LL} \to
 \delta_{RR}}, \label{up squark}
\end{eqnarray}
where $M_{Q_i}$ and $M_{U_i}$ ($i=1,2,3$ denotes generation index) are soft breaking parameters with mass dimension, $M^{U}_{LR}$ and $M^{U}_{RL}$ represents
SUSY scale defined as $M^{U}_{LR} = (M_{U_3} + M_{Q_2})/2$ and $M^{U}_{RL} = (M_{U_2} + M_{Q_3})/2$, and $\delta_{LL}$, $\delta_{LR}$, $\delta_{RL}$ and $\delta_{RR}$ reflect the extent of the flavor violation.
Similarly, for down-squarks we have
\begin{eqnarray}
(M^2_{\tilde{d}})_{LR}& = & \left ( \begin{array}{ccc}
0          &  0                  &  0 \\
0                                &  0   & \delta_{LR}^d v_1 M^{D}_{LR}\\
0 &  \delta_{RL}^d v_1 M^D_{RL}  & m_b A_b \end{array}  \right
),  \nonumber  \\
(M^2_{\tilde{d}})_{RR} &= & (M^2_{\tilde{u}})_{LL}|_{M_{Q_i}^2 \to
M_{D_i}^2,~ \delta_{LL} \to \delta_{RR}^d}, \label{delta'-LR}
\end{eqnarray}
and due to $SU(2)$ gauge symmetry, $(M^2_{\tilde d })_{LL}$
is determined by\cite{susyf}
\begin{eqnarray}
(M^2_{\tilde{d}})_{LL} = V_{CKM}^\dag (M_{\tilde{u}}^2)_{LL}
V_{CKM}  \label{SU2}
\end{eqnarray}
with $V_{CKM}$ denoting the Cabibbo-Kobayashi-Maskawa matrix in the SM.
Note that in Eqs.(\ref{up squark},\ref{delta'-LR}), we only keep
the chiral-flipping terms for third-family squarks because these terms are
usually assumed to be proportional to corresponding quark masses, and
can not be neglected only for  third-family squarks.

The squark mass eigenstates can be obtained by diagonalizing the
mass matrix presented above with an unitary rotation $U_{\tilde{q}}$, which
is performed numerically in our analysis. The interaction of the field $X$ with a pair of
squark mass eigenstates is then obtained by
\begin{eqnarray} V(X \tilde{q}_{\alpha}^\ast
\tilde{q}_{\beta}^\prime) \; = \;
 U^{\dag \tilde{q}}_{\alpha,i} \;  U^{\tilde{q}^\prime}_{j, \beta} \;
 V(X \tilde{q}_i^\ast \tilde{q}^\prime_j)~,
\label{rotation}
\end{eqnarray}
where $V(X\tilde{q}_i^\ast \tilde{q}^\prime_j)$ denotes a generic
vertex in the interaction basis and
$V(X\tilde{q}_{\alpha}^\ast \tilde{q}_{\beta}^\prime )$ is the
vertex in the mass-eigenstate basis. It is clear that both the
squark masses and their interactions depend on the mixing
parameters $\delta_is$.

In some fundamental supersymmetric theories like the mSUGRA and gauge-mediated
SUSY-breaking models, the mixing parameters are functions of the soft breaking masses
and usually exhibit certain hierarchy structure\cite{flavor-mixing}. In this work, in order to
make our discussion as general as possible, we treat all $\delta_is$
as free parameter, and limit them by some physical observables.
The constraints we consider include
\begin{itemize}
\item[(I)]The recently measured SM-like Higgs boson mass $m_h$. In the MSSM, this mass is determined by
the renormalized self-energies of the doublet CP-even  Higgs fields, $h_u$ and $h_d$, and the transition
between them. Squarks contribute to these quantities through the $\tilde{q}^\ast \tilde{q} S$ and
$\tilde{q}^\ast \tilde{q} S S$ interactions  with $S$ denoting either $h_u$ or $h_d$\cite{NMF-Higgsmass,cao-flavor-mixing}.
In the presence of the flavor mixings, both the interactions and the squark masses
may be quite different from those in the case of $\delta_is=0$, and so is the SM-like Higgs boson mass.
Among the flavor mixing parameters $\delta_i$, the Higgs mass is more sensitive to the chiral-flipping ones
$\delta_{LR}$ and $\delta_{RL}$.

In this work, we get $m_h$ in the MSSM by the code FeynHiggs\cite{FeynHiggs}. In our scan over the parameter space
of the low energy MSSM, we require the mass to be about $2 {\rm GeV}$ around its measured central value, i.e.
$123 {\rm GeV} \leq m_h \leq 127 {\rm GeV}$. While for the MSSM in heavy SUSY case (see below),
noting that the mass obtained by FeynHiggs suffers from potentially large theoretical uncertainties,
we require a moderately wider range,  i.e. $121 {\rm GeV} \leq m_h \leq 129 {\rm GeV}$.

\item[(II)] The LHC search for SUSY. By now both the ATLAS and CMS collaborations have paid great effects in searching for
the signals of gluino, squarks as well as charginos and neutralinos,
and based on certain assumptions, they exclude some SUSY particles up to about 1 TeV\cite{SUSY-search}.
These obtained results, however, can not be applied directly to a general SUSY case, and in order to implement the LHC constraints, one
has to perform detailed Monte Carlo simulation for each SUSY parameter point with the same strategies as
those of the collaborations, then compare the simulated results with the LHC data\cite{Natural-SUSY-search}.
In practice, such a process is rather time consuming, and can not be applied to an extensive scan over the SUSY
parameter space, where a large number of samples are involved.

In order to simplify our analysis, we note that by now gluino is preferred to be at TeV scale
without considering special cases such as compressed SUSY spectra\cite{SUSY-search}, while the second and third generation squark may
still be as light as several hundred GeV\cite{Natural-SUSY-search,Light-second-generation}, especially in the presence of the flavor mixing when the
limitation on the squark spectrum can be further relaxed\cite{Flavored-natural}. So we make following
assumption in our discussion
\begin{eqnarray}
m_{\tilde{q}_\alpha},  m_{U_3} \geq 200 {\rm GeV}, \quad  m_{\tilde{g}} \geq 1 {\rm TeV}, \quad m_{Q_{2}}, m_{U_2},
m_{Q_3} \geq 500 {\rm GeV}.
\end{eqnarray}
As will be shown below, our conclusions are not sensitive to such assumption.

\item[(III)] The metastability of the vacuum state. This constraint reflects the fact that
squarks as scalar fields contribute to SUSY potential, and consequently their soft breaking
parameters should be limited by the stability (or more general metastability) of the vacuum state\cite{metastability1,metastability2}.
Assuming only $\delta_{LR}$ or $\delta_{RL}$ contributes to the potential, the metastability requires\cite{metastability1,metastability2}
\begin{eqnarray}
|\delta_{LR}^u| &\lesssim & 1.2\times \frac{m_t}{v_2} \frac{\sqrt{M_{Q_2}^2 + M_{U_3}^2 + M_A^2 \cos^2 \beta}}{M^{U}_{LR}}, \nonumber \\
|\delta_{RL}^u| &\lesssim & 1.2\times \frac{m_t}{v_2} \frac{\sqrt{M_{U_2}^2 + M_{Q_3}^2 + M_A^2 \cos^2 \beta}}{M^{U}_{RL}}, \nonumber \\
|A_t| & \lesssim & 2.67 \sqrt{M_{Q_3}^2 + M_{U_3}^2 + M_A^2 \cos^2 \beta}.
\end{eqnarray}
Note in previous study on top quark flavor changing neutral current process, this constraint was usually missed.
\item[(IV)] EW precision observables $M_W$ and $\sin^2 \theta_{eff}$. In SUSY, the corrections to $M_W$ and $\sin^2 \theta_{eff}$
are dominated by squark loops, and their sizes reflect the mass disparity of left-handed SU(2) doublet squarks.
To a good approximation, these two quantities are related with the $\delta \rho$ parameter by
    \begin{eqnarray}
\delta M_W \simeq \frac{M_W}{2} \frac{c_W^2}{c_W^2 -s_W^2} \delta \rho, \quad \delta \sin^2 \theta_{eff} \simeq -\frac{c_W^2 s_W^2}{c_W^2
-s_W^2} \delta \rho,
\end{eqnarray}
where
\begin{eqnarray}
 \delta \rho=\frac{\Sigma_Z(0)}{M_Z^2} - \frac{\Sigma_W(0)}{M_W^2}.
\end{eqnarray}
In this work, we repeat our previous calculation of $\delta M_W$ and
$\delta \sin^2 \theta_{eff}$ in \cite{cao-flavor-mixing} where three generation squarks are considered to
implement the $SU(2)$ relation between $(M_{\tilde{U}}^2)_{LL}$ and $(M_{\tilde{D}}^2)_{LL}$, and require
\begin{eqnarray}
\delta M_W \leq 21 MeV, \quad \quad \delta \sin^2 \theta_{eff} \leq 19.6\times 10^{-5},
\end{eqnarray}
which are their allowed ranges at $2\sigma$ level after considering experimental and
theoretical uncertainties\cite{EW-observable}.

\item[(V)] Constraint from $B \to X_s \gamma$.  In the MSSM, the SUSY contributions to $B\to X_s \gamma$ come from four kinds of
loops mediated by charged Higgs bosons, charginos, neutralinos and gluinos respectively. We calculate these contributions by the code
FeynHiggs\cite{FeynHiggs}, and require $3.04\times 10^{-4} \leq Br(B \to X_s \gamma) \leq 4.02\times 10^{-4} $ (corresponding its $2\sigma$ allowed range
by experiments\cite{bsg}) in our parameter scan.

Since the neutralino contribution is usually small, the expression of $B \to X_s \gamma$ in
the NMSSM is roughly identical to that of the MSSM.
\end{itemize}

About above constraints, it should be noted that constraints (I), (III) and (IV) do not diminish as SUSY scale becomes higher, that is,
they are not decoupled in heavy SUSY limit; while the process $ B\to X_s \gamma$ does not possess such a property.

\begin{figure}[t]
\includegraphics[width=12cm]{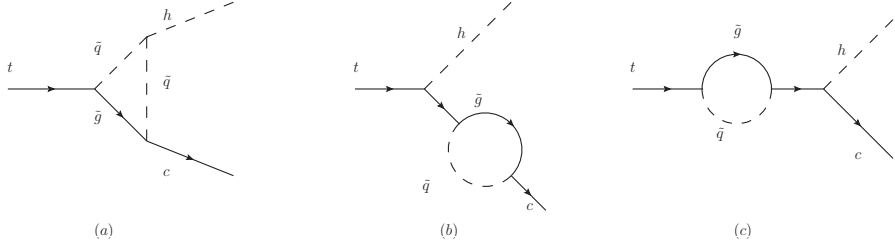}
\caption{Feynman diagrams of the SUSY-QCD contribution to $t \to ch$. If charm quark mass is neglected,
the contribution from diagram (c) vanishes.}
\label{fig1}
\end{figure}

\section{SUSY prediction on the rate of $t \to c h$ }

In the MSSM, the dominant contribution to the process $t \to ch$ arises from the SUSY-QCD diagrams
shown in Fig.\ref{fig1}.  The relevant Lagrangian is given by\cite{Mixing-FeynRule}
\begin{eqnarray}
{\cal{L}} &=&  \sqrt{2}g_{s} [ \bar{\tilde{g}}_{a} (-U^{\ast}_{2 \alpha}P_{L} + U^{\ast}_{5 \alpha}P_{R}) \tilde{q}_{\alpha i}^{\ast} T_{i j}^{a} c_{j}
+  \bar{\tilde{g}}_{a} (-U^{\ast}_{3 \alpha}P_{L} + U^{\ast}_{6 \alpha}P_{R}) \tilde{q}_{\alpha i}^{\ast} T_{i j}^{a} t_{j} ] + h.c.  \nonumber \\
& & + \sum_{\alpha,\beta=1}^6 C_{\alpha \beta} \tilde{q}^\ast_{\alpha} \tilde{q}_\beta h  + Y_t \bar{t} t h,
\label{lagrangian}
\end{eqnarray}
where $\tilde{g}$ and $\tilde{q}_\alpha$ denote gluino and squark in the mass eigenstate respectively,
$T^{a}$ is the Gell-Mann matrix with $i$ and $j$ representing color indices, $U_{\tilde{q}}$ is
the $6\times6$ rotation matrix to diagonalize the mass matrix for up-type squarks, $C_{\alpha \beta}$
parameterizes the coupling of the Higgs boson with squark mass eigenstates $\tilde{q}_\alpha$ and $\tilde{q}_\beta$,
and $Y_t = \frac{m_t}{v}\frac{\cos \alpha}{\sin \beta}$ with $\alpha$ being the rotation angle to
diagonalize the CP-even Higgs mass matrix. The amplitude of $t \rightarrow c h $ can then be expressed by
\begin{equation}
i \mathcal{M}(t \rightarrow c h) = \bar{u}(p_{c})[(F_{1L}+F_{2L})P_{L} + (F_{1R}+F_{2R})P_{R}]u(p_{t}),
\end{equation}
where, after neglecting charm quark mass, $F_i$s are given by
\begin{eqnarray}
 F_{1L} &=& \frac{i g^{2}_{s}}{6{\pi}^{2}}\{
 C_{\alpha\beta}U^{\ast}_{3 \alpha}U_{5 \beta}m_{\tilde{g}} C_0(p_c^2,p_h^2,p_t^2,m_{\tilde{g}}^2,m^{2}_{{\tilde{q}}_{\alpha}},m^{2}_{{\tilde{q}}_{\beta}})\nonumber \\
 & &-
 C_{\alpha\beta}U^{\ast}_{6 \alpha}U_{5 \beta}m_t C_{12}(p_c^2,p_h^2,p_t^2,m_{\tilde{g}}^2,m^{2}_{{\tilde{q}}_{\alpha}},m^{2}_{{\tilde{q}}_{\beta}})
 \}, \nonumber \\
    F_{1R} &=& \frac{i g^{2}_{s}}{6{\pi}^{2}}\{
 C_{\alpha\beta}U^{\ast}_{6 \alpha}U_{2 \beta}m_{\tilde{g}} C_0(p_c^2,p_h^2,p_t^2,m_{\tilde{g}}^2,m^{2}_{{\tilde{q}}_{\alpha}},m^{2}_{{\tilde{q}}_{\beta}})\nonumber \\
 & &-
 C_{\alpha\beta}U^{\ast}_{3 \alpha}U_{2 \beta}m_t C_{12}(p_c^2,p_h^2,p_t^2,m_{\tilde{g}}^2,m^{2}_{{\tilde{q}}_{\alpha}},m^{2}_{{\tilde{q}}_{\beta}})
 \}, \nonumber \\                                                       F_{2L} &=& -\frac{i g^{2}_{s}m_{\tilde{g}}}{6{\pi}^{2}m_{t}}Y_{t}U^{\ast}_{3 \alpha}U_{5 \alpha} B_0(p_c^2,m_{\tilde{g}}^2,m^{2}_{{\tilde{q}}_{\alpha}}), \nonumber \\
 F_{2R} &=& -\frac{i g^{2}_{s}m_{\tilde{g}}}{6{\pi}^{2}m_{t}}Y_{t}U^{\ast}_{6 \alpha}U_{2 \alpha} B_0(p_c^2,M_{\tilde{g}}^2,m^{2}_{{\tilde{q}}_{\alpha}}).
\end{eqnarray}
In above expressions, $p_t$, $p_c$ and $p_h$ denote the momentums of top quark, charm quark, and Higgs boson respectively,
$m_{\tilde{g}}$ and $m_{{\tilde{q}}_{\alpha}}$ represent the masses of gluino and squark respectively,
and $B_0$ , $C_0$ and $C_{12}$ are the standard two-point and three-point loop functions respectively\cite{Loop-function}.
In the heavy SUSY case discussed below, since the involved sparticle masses are much larger than $m_t$,
the contribution from $C_{12}$ can be safely ignored, and $B_0$, $C_0$ can be approximated by
\begin{eqnarray}
C_0(p_c^2,p_h^2,p_t^2,m_{\tilde{g}}^2, m^{2}_{{\tilde{q}}_{\alpha}}, m^{2}_{{\tilde{q}}_{\beta}}      )  &=&
\frac{1}{m_{\tilde{g}}^2}  \frac{1}{1-{\delta}_{\beta}} [\frac{{\delta}_{\beta}}{{\delta}_{\alpha}-{\delta}_{\beta}}\ln(\frac{{\delta}_{\alpha}}{{\delta}_{\beta}})
-\frac{1}{{\delta}_{\alpha}-1}\ln{\delta}_{\alpha}  ] + {\cal{O}}(\frac{p_t^2}{m_{\tilde{g}}^2}), \nonumber  \\
B_0(p_c^2,m_{\tilde{g}}^2,m^{2}_{{\tilde{q}}_{\alpha}}) &=& 1 + \frac{{\delta}_{\alpha}}{1-{\delta}_{\alpha}}
\ln{\delta}_{\alpha} +{\cal{O}}(\frac{p_c^2}{m_{\tilde{g}}^2}),\label{code}
\end{eqnarray}
where $ {\delta}_{\alpha}$ is defined as ${\delta}_{\alpha}= m^{2}_{{\tilde{q}}_{\alpha}}/m^{2}_{\tilde{g}}$.

\subsection{$t \to c h$ in low energy SUSY}

As shown in \cite{cao-top-fcnc}, the SUSY-QCD contribution to $t \to c h$ in low energy MSSM has following features
\begin{itemize}
\item In case that only one flavor mixing parameter $\delta_i$ is non-zero, the rate of $t \to c h$ increases
monotonously as the $\delta_i$ becomes larger, while if several non-vanishing $\delta_i$s coexist, their effects
may cancel each other out.
\item Since the effective $h\bar{c}t$ vertex involves both chiral-flipping and flavor-changing,
the chiral-conserving parameter $\delta_{LL}$/$\delta_{RR}$ must be accompanied with chiral-flipping
$h\tilde{t}_L^\ast \tilde{t}_R$ or $h\tilde{t}_R^\ast \tilde{t}_L$ interaction in contributing to the vertex, while the chiral-flipping parameter
$\delta_{LR}$/$\delta_{RL}$ alone is able to lead into the $h\bar{c}t$ vertex. As a result, the effective vertex
is usually more sensitive to $\delta_{LR}$ and $\delta_{RL}$ if we do not consider the constraints on the mixing parameters.

\item Unlike the other top quark FCNC processes, the Super-GIM mechanism  does not apply to the decay $t\to c h$. But since there
 exists a strong cancelation between diagram (a) and (b) in Fig.\ref{fig1} (see discussion below), the rate of $t \to c h$ depends on the soft mass parameters in a complex way.

\end{itemize}

\begin{table}
\caption{Benchmark points in low energy SUSY which correspond to very optimal cases in predicting
the rate of $ t \to c h$. Points 1 and 2 are obtained from Scan-I, and Points 3 and 4 are from
Scan-II. All these points satisfy the constraints listed in the text. \label{table1}}

\vspace{0.2cm}

\begin{tabular}{|c|c|c|c|c|}
\hline
             & Point 1 & Point 2 & Point 3 & Point 4 \\\hline
$M_{Q2}$       & 1694GeV & 945GeV  & 528GeV  & 516GeV \\\hline
$M_{Q3}$       & 551GeV  & 519GeV  & 1704GeV & 1672GeV\\\hline
$M_{U2}$       & 1901GeV & 1633GeV & 759GeV  & 704GeV \\\hline
$M_{U3}$       & 988GeV  & 1013GeV & 1659GeV & 1770GeV\\\hline
$A_t$          &-1110GeV &-1249GeV &-3287GeV & 3150GeV\\\hline
$\delta_{LL}$  & -0.0600 &-0.0956  & -0.4934 & 0.2720 \\\hline
$\delta_{LR}$  & -0.8447 & 0.0548  & 0.4710  & 1.108  \\\hline
$\delta_{RL}$  & 0.9154  & 1.265   &-1.583   &-1.560  \\\hline
$\delta_{RR}$  & 0.7523  & 0.6198  & -0.7913 & 0.7959 \\\hline
$\tan\beta$    & 35.5    & 39.8    & 1.5     & 1.5    \\\hline
$M_{\tilde{g}}$  & 1128GeV & 1291GeV & 1037GeV & 1067GeV\\\hline
$m_A$          & 1151GeV & 1393GeV & 1224GeV & 906GeV\\\hline
$\mu$          & 1958GeV & 1848GeV & 832GeV  & -973GeV\\\hline
$\lambda$      & $-$     & $-$     & 0.690   & 0.699  \\\hline
$Br(t \to ch)$ & $2.99\times10^{-6}$ & $2.82\times10^{-6}$ & $1.15\times10^{-5}$ & $0.90\times10^{-5}$\\\hline
\end{tabular}
\end{table}

In the following, we do not intend to exhibit these features, but instead, noting that
the constraints (I-III) introduced above were not considered before we try to figure out the order of
$Br(t \to c h)$ that SUSY can predict after considering these constraints. For this purpose,
we perform two independent scans over relevant SUSY parameters by imposing the constraints. Details of our scans are as follows:
\begin{itemize}
\item Scan-I: We restrict our discussion in the MSSM, and calculate the Higgs mass with the code FeynHiggs.
The parameter region we explore is given by
 \begin{eqnarray}
 && 500 {\rm GeV} \leq m_{Q_2}, m_{Q_3},m_{U_2}, m_{D_2}, m_{D_3} \leq 2 {\rm TeV}, \quad 200 {\rm GeV} \leq m_{U_3} \leq 2 {\rm TeV}, \nonumber \\
 && |A_t| \leq 6  \sqrt{m_{Q_3} m_{U_3}}, \quad  1 TeV \leq  m_{\tilde{g}}  \leq 2 {\rm TeV}, \quad 1 \leq \tan\beta \leq 40, \nonumber \\
 &&  -1 \leq \delta_{LL},\delta_{RR} \leq 1, \quad
 -2.0\leq \delta_{LR},\delta_{RL} \leq 2.0, \quad -0.5 \leq \delta_{LR}^d,\delta_{RL}^d \leq 0.5,  \nonumber \\
 &&400 {\rm GeV} \leq m_A \leq 2 {\rm TeV}, -2 {\rm TeV} \leq \mu \leq 2 {\rm TeV}.
\end{eqnarray}
In drawing up the strategy of this scan, we note that, although the down type squark parameters like $\delta_{LR}^d$ and $M_{D}$
do not affect the rate of $t \to c h$, they are  needed in $\delta \rho$ and $B \to X_s \gamma $ calculation. So to make our conclusions as general
as possible we vary them in reasonable regions. We also note that since too many parameters are involved in the scan, the traditional random scan method
is not efficient in searching for maximal value of $Br( t \to c h)$. So we adopt the Markov chain method in doing such a job. During the scan
we adjust the optimal value by the results obtained from previous samplings until it reaches some stable values.

\item Scan-II: Same as Scan-I, but in order to relax the Higgs mass bound, we go beyond the MSSM by considering extra
contribution to the mass. To be more specific, now we write the Higgs mass as $m_h^2 =
m_{h,MSSM}^2 + \lambda^2 v^2 \sin^2 2\beta$ with $\lambda$ being a free parameter in the range from 0 to 0.7.
This treatment of the Higgs mass is motivated by the singlet extensions of the MSSM such as the NMSSM,
where the interaction between the singlet Higgs field and the doublet Higgs fields in the MSSM provides
an additional contribution to the mass at tree level\cite{NMSSM}.
In order to get a large contribution from this singlet extension, we set $\tan\beta =1.5$.

\end{itemize}

In Table \ref{table1}, we show two benchmark points for each scan. These points correspond to very optimal
cases in predicting a large rate of $t \to c h$. After analyzing our scan results, we have following observations
\begin{itemize}
\item A small $m_{\tilde{g}}$ (around its experimental lower bound) seems to be favored to
maximize the rate of $t \to c h$ after considering the constraints. Meanwhile, it is interesting to learn from
Table \ref{table1} that, although we allow $m_{U_3}$ to be as low as $200 {\rm GeV}$,  the optimal points do not correspond to low $m_{U_3}$s.
\item In low energy SUSY, the size of $Br( t \to c h)$ for the optimal points in Table \ref{table1} is not sensitive to the parameters $m_A$ and $\mu$.
For example, our results indicate that shifting $m_A$ from its value in Table \ref{table1} by 100 GeV only results in a change of $Br( t \to c h)$ by less than $1\%$.
While as will be shown below,  these two parameters play an important role in determining the rate of the rare decay in heavy SUSY limit.
\item   Among the considered constraints, the most stringent ones come from the Higgs mass and the LHC search for SUSY. As a result,
the branching ratio of $t \to c h$ can only reach $10^{-6}$ in low energy MSSM, which is about one order smaller than previous
predictions obtained before the advent of the LHC\cite{cao-top-fcnc}. In scan-II, however, since the Higgs mass constraint on squark masses is 
comparatively relaxed, $Br(t \to c h)$ can reach $10^{-5}$.
\end{itemize}

About our results on $t \to c h$, we remind that we do not include the SUSY-EW contribution to
$t \to c h$. The reason is the amplitude of the SUSY-EW contribution is roughly determined by
$\alpha m_{\tilde{\chi}^\pm}/Max(m_{\tilde{q}_{D}}^2$, $m_{\tilde{\chi}^\pm}^2)$ from naive estimation\cite{SUSYEW-tch}, while the SUSY-QCD
contribution is determined by $\alpha_s m_{\tilde{g}}/Max(m_{\tilde{q}_{U}}^2$, $m_{\tilde{g}}^2)$,  where
$m_{\tilde{q}_{D}}$, $m_{\tilde{q}_{U}}$ and $m_{\tilde{\chi}^\pm}$ denote the mass scales for down-type squarks,
up-type squarks and charginos respectively. Noting that $m_{\tilde{q}_{D}}$ is not much smaller than $m_{\tilde{g}}$
as suggested by the LHC search for the second and third generation squarks\cite{Light-second-generation,SUSY-search},
and also that the flavor mixings in the down-type squark sector are more tightly constrained by B-physics than those in
up-type squark sector, we conclude that the SUSY-EW contribution should not be comparable
with the SUSY-QCD contribution. So our estimates on the magnitude of $t \to c h$ will not change
after including the SUSY-EW contribution. Another reason to neglect the EW contribution is, once considering it,
too many parameters will be involved, but meanwhile this does not change our conclusion.

Before we end this subsection,  we have two comments. One is in extensions of the MSSM, the decay chain of a certain sparticle may be quite
different from its MSSM prediction, and the analysis of the ATLAS and CMS collaborations in searching for SUSY
may become irrelevant\cite{Ulrich-Singlino}. As a result, the constraint on the sparticle mass may be relaxed. This in return may push up the SUSY
prediction on the rate of $ t \to c h$. Moreover, in extensions of the MSSM the couplings of the Higgs boson with quarks and squarks may be slightly changed. The influence of such changes on the rate of $ t \to c h$ is usually not as signficant as the relax of the Higgs mass constraint.
The other comment is that, given $Br(t \to c h) \sim 10^{-5}$, it is difficult to detect such a top quark
rare decay
at the 14-TeV LHC with an integrated luminosity of $100 fb^{-1}$\cite{Top-Workinggroup}, but at future linear colliders such as
TLEP\cite{TLEP}, detection of the decay is still possible.

\subsection{Remanent effect of $t \to c h$ in heavy SUSY limit}

\begin{figure}[t]
   \begin{center}
        \begin{tabular}{cc}
            \includegraphics[width=85mm]{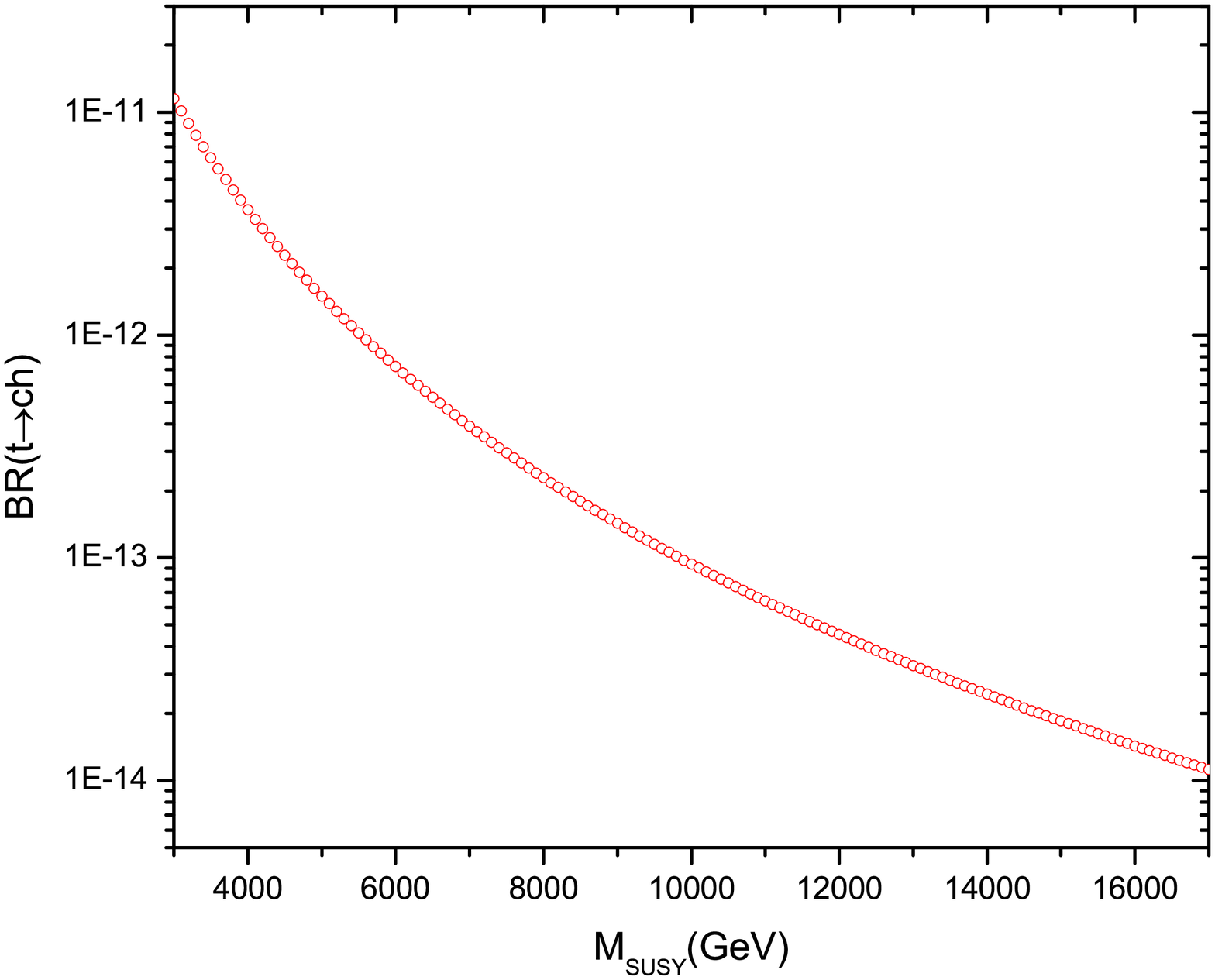} &
            \includegraphics[width=85mm]{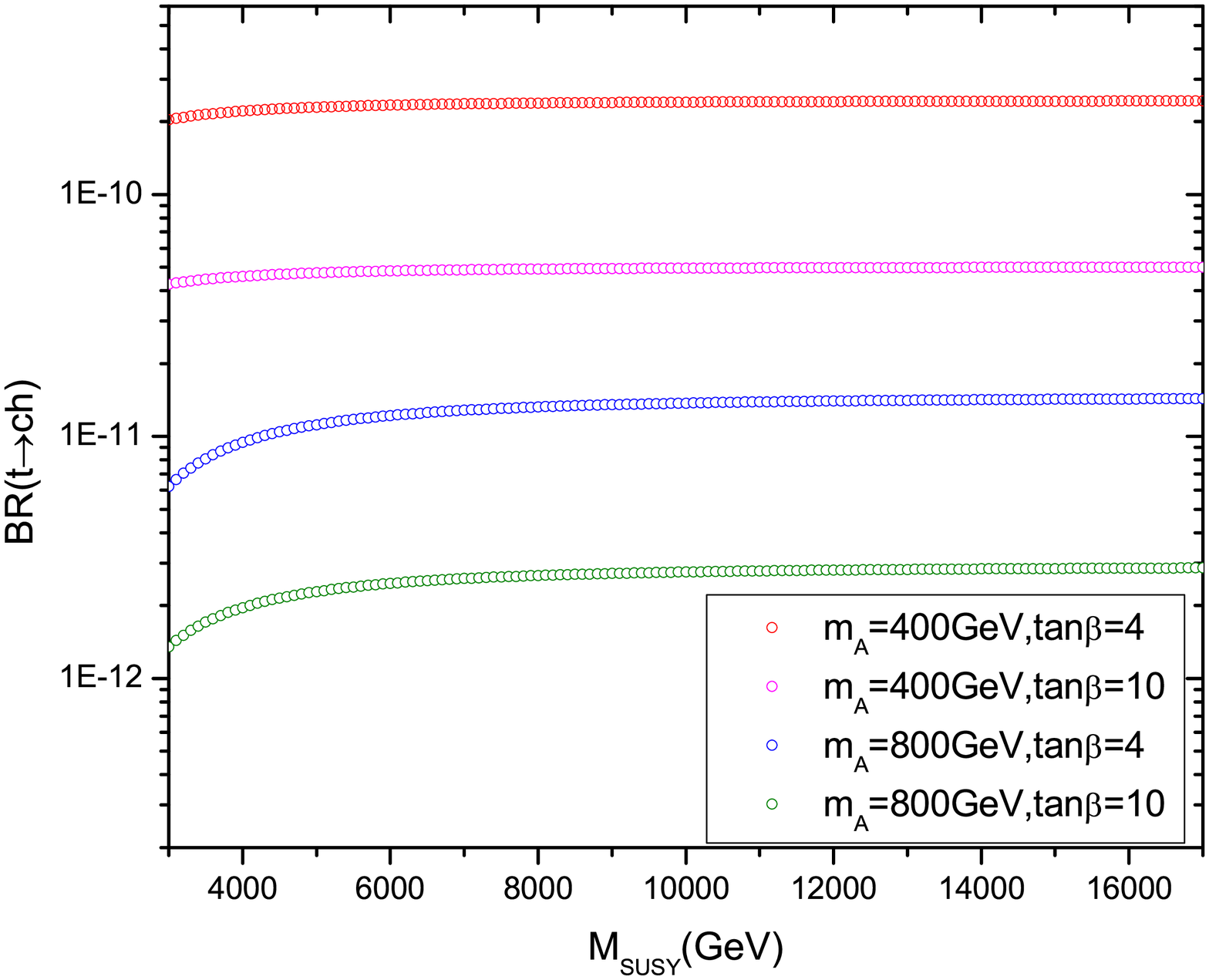}
        \end{tabular}
   \end{center}
\caption{Dependence of $Br(t \to c h)$ on the common squark mass scale $M_{SUSY}$. In getting this figure, we assume
all soft masses equal to $M_{SUSY}$, $\delta_{LL}=0.7$, $\tan \beta = 4,10$ and $m_A = 400,800GeV$.
We set $A_t = M_{SUSY}$ and $\mu = 0$ for the left panel, and $A_t = 0$ and $\mu = M_{SUSY}$ for the right panel.
Note the lines corresponding to different choices of $ m_A $ and $\tan \beta$ overlap in left panel.}
\label{fig2}
\end{figure}

As mentioned in Section I, because the $h\tilde{q}^\ast \tilde{q}$ coupling strength is mainly determined by soft SUSY breaking parameters,
the SUSY-QCD contribution to the effective $h\bar{c}t$ interaction may exhibit remanent effect of SUSY  in heavy SUSY limit.
In order to investigate such an issue,
we in the following assume a common SUSY mass scale $m_{Q_2} = m_{Q_3} = m_{U_2} = m_{U_3} = m_{\tilde{g}}=M_{SUSY}$, and study the
dependence of $Br(t \to c h)$ on $M_{SUSY}$ for different choices of $A_t$ and $\mu$ in Fig.\ref{fig2}.
In getting Fig.\ref{fig2}, we consider the case that only $\delta_{LL}$ is non-vanishing and fix $\delta_{LL} = 0.7$,
$\tan \beta = 4, 10$ and $m_A = 400, 800 {\rm GeV}$.  We set $A_t = M_{SUSY}$ and $\mu = 0$ for left panel,  and $A_t = 0$ and
$\mu = M_{SUSY}$ for right panel. These settings are only for exhibiting the decoupling behavior of $Br(t \to c h)$ and we 
do not consider the constraint from the LEP search for charginos,
which requires $\mu \gtrsim 103 {\rm GeV}$. Fig.\ref{fig2} then indicates that SUSY has remanent effect on the rare decay rate
only when $\mu$ is at SUSY scale and meanwhile $m_A$ is
at weak scale, and the size of the effect depends strongly on $m_A$ and $\tan \beta$, e.g. small values of $m_A$ and $\tan \beta$ tend to
enhance the effect.  We also investigate the case that only $\delta_{LR}$ and $\delta_{RL}$ are non-vanishing, and we do not find
such remanent effect in heavy SUSY limit for any choices of $\mu$ and $m_A$.

The behaviors shown in Fig.\ref{fig2} can be understood by the effective Lagrangian that
describes the Higgs and quark system. After including loop effects, one can write down the Lagrangian as follows
\begin{eqnarray}
{\cal{L}}&=&  \sum_{i,j=1}^3 \{ \bar{q}^\prime_i ( m^\prime_{ij} + \delta m^\prime_{ij}) P_L q^\prime_j + h \bar{q}^\prime_i ( Y^\prime_{ij} + \delta Y^\prime_{ij}) P_L q^\prime_j \} + h.c. \nonumber \\
         &=&  \sum_{i,j=1}^3 \{ (\bar{q} V_R^\dag)_i ( m^\prime_{ij} + \delta m^\prime_{ij}) P_L ( V_L q )_j + h (\bar{q} V_R^\dag)_i ( Y^\prime_{ij} + \delta Y^\prime_{ij}) P_L ( V_L q )_j \} + h.c. \nonumber \\
         &=& \sum_{i=1}^3 \{ \bar{q}_i m_i P_L q_i + \frac{h}{v} \bar{q}_i m_i P_L q_i \}
         + \sum_{i,j=1}^3 h \bar{q}_i [V_R^+ (\delta Y^\prime - \frac{\delta m^\prime}{v} ) V_L]_{ij} P_L q_j + h.c.,
\end{eqnarray}
where $q^\prime_i$ and $m_{ij}^\prime = Y^\prime_{ij} v$ are quark field and its mass matrix at tree level with $i,j$ denoting flavor indices, and $\delta m^\prime$ and $\delta Y^\prime$ represent loop corrections to the mass matrix and the Yukawa coupling respectively. The second equation reflects the definition of quark mass eigenstate with $V_L$ and $V_R$ denoting the rotation matrices for left-handed quarks and
right-handed quarks respectively. After such a definition, the loop corrected mass matrix $m^\prime + \delta m^\prime$
is diagonal with its diagonal element $m_i$ representing physical quark mass determined by experiments. At this stage,
the correction to the $h\bar{q}_i q_j$ interaction is given by $h \bar{q}_i [V_R (\delta Y^\prime - \delta m^\prime/v ) V_L]_{ij} P_L q_j + h.c.$.
Obviously, if $\delta Y^\prime = \delta m^\prime/v$, new physics contribution to the $h\bar{q}_i q_j$ interaction vanishes.
In actual calculation, $\delta m^\prime$ is obtained from $q_i-q_j$ transition diagrams like diagram (b) of Fig.\ref{fig1}
without the emission of the Higgs particle, and $\delta Y^\prime$ comes from the vertex correction like diagram (a) of
Fig.\ref{fig1}. The effective Lagrangian then indicates that the two contributions should cancel out each other
in contributing to the $h\bar{q}_i q_j$ interaction.

As far as the SUSY-QCD correction to the $h\bar{q}_i q_j$ interaction is concerned, its behavior in heavy SUSY limit
can be analysised with the mass insertion approximation. In this method, squark masses are taken to be the
diagonal elements of the squark mass matrix, and the non-diagonal elements are treated as interactions. In order
to illustrate this method in explaining the remanent effect, we first consider the well studied SUSY-QCD
correction to the $h\bar{b}b$ vertex\cite{SUSY-QCD-hbb}. In this example, to get $\delta m_b$ one needs to insert the $\tilde{b}_L-\tilde{b}_R$
transition  by odd times into the sbottom propagator entered in bottom quark self-energy diagrams, and sum the
corresponding contribution to infinite orders of the insertion. One can check that, with one more insertion,
the corresponding contribution is suppressed by a factor $m_b(A_b - \mu \tan \beta)/M_{SUSY}^2$ compared with that without the insertion,
and only for the first order insertion, $\delta m_b$ is not suppressed by $M_{SUSY}$, which means that the corresponding
contribution is non-decoupled in heavy SUSY limit. In a similar way, one can check that even times of the insertion are
needed to get the expression of $\delta Y_b$ in calculating the $h\bar{b}b$ vertex correction, and only for the
zero-th insertion, the contribution is non-decoupled. Putting these two contributions together, one can learn that
the non-decoupled terms proportional to $A_b$ is exactly canceled out, and the remaining contribution is proportional to the well known
form $\mu m_{\tilde{g}}/M_{SUSY}^2 (\tan \beta + \cot \alpha) \simeq  (-2 m_Z^2 \mu m_{\tilde{g}})/(M_{SUSY}^2 m_A^2)
\tan \beta \cos 2 \beta$\cite{SUSY-QCD-hbb}.  This effect, albeit scaling as $1/m_A^2$, does not diminish as $\mu \simeq m_{\tilde{g}} \simeq 
M_{SUSY}$ approaches infinity for $m_A$ at weak scale, and is therefore dubbed as the remanent effect of SUSY.

Next we turn to analysis the SUSY-QCD contribution to the $h\bar{c}t$ vertex. Since such a interaction involves both
chiral-flipping and flavor-changing, appropriate insertions are needed to accomplish both the tasks. For the
chiral-flipping mixings $\delta_{LR}$ and $\delta_{RL}$, their role is quite similar to $A_b$ in the SUSY-QCD
correction to the $h\bar{b}b$ coupling, and their non-decoupling contribution is completely canceled out. While
for the chiral-conserving mixings $\delta_{LL}$ and $\delta_{RR}$, their contribution to the $h\bar{b}b$ vertex can be split
into two parts with one part proportional to $A_t$ and the other part proportional to $\mu$. Fig.\ref{fig2} then reflects
that the non-decoupling contribution of the former part is exactly canceled out,
while that of the latter part is maintained if $m_A$ is not at the same order as $M_{SUSY}$. This situation is actually similar
to the SUSY-QCD correction to $h\bar{b}b$ vertex with the only difference coming from the fact that such remanent effect is not enhanced by
$\tan \beta$.  In fact, if both $M_{SUSY}$ and $m_A$ approach infinity simultaneously,
the genuine SUSY contribution to $t\to c h$ should vanish since now the Higgs sector of the MSSM is identical to
that of the SM, while if only  $M_{SUSY}$ approaches infinity, the Higgs sector is described by a Two-Higgs-Doublet model,
and SUSY may leave its imprint in Higgs sector\cite{more-discuss}. Our results in Fig.\ref{fig2} actually reflect such a possibility.

\begin{table}[t]
\caption{Same as Table \ref{table1}, but for heavy SUSY case. Point 5 and Point 6 are taken
from Scan-III and Scan-IV respectively. \label{table2}}

\vspace{0.2cm}

\begin{tabular}{|c|c|c|c|c|c|c|c|c|}
\hline
 &$M_{Q2} ({\rm GeV}) $ &$M_{Q3}({\rm GeV}) $ &$M_{U2} ({\rm GeV})$ &$M_{U3} ({\rm GeV})$ & $A_t ({\rm GeV})$  &  $\delta_{LL}$ & $\delta_{LR}$ & $\delta_{RL}$\\ \hline
Point 5 &  4484 &  4039 &  6871 & 6839  &  6878 & -0.4885   &1.202   &-1.208    \\ \hline
Point 6 & 4158 & 4046 & 6886 & 6954 & 3944 & -0.3382  & -0.1845   & -0.7258  \\ \hline
& $\delta_{RR}$ & $\tan \beta$ & $m_{\tilde{g}} ({\rm GeV})$ & $m_A ({\rm GeV})$ & $\mu ({\rm GeV})$ & $\lambda$ & $Br(t \to ch)$& \\ \hline
Point 5 & 0.7352   &  6.13     & 4095 &800 & -5943 & $-$ & $1.04\times10^{-8}$& \\ \hline
Point 6 & 0.7734  &1.5      &  4138 & 800 & -18320 & 0.690  &$1.65\times10^{-8}$& \\ \hline
\end{tabular}
\end{table}

We finally discuss how large SUSY can predict the rate of $t \to c h$ in heavy SUSY case.
For this purpose, we require all squarks to be heavier than $3 {\rm TeV}$ and perform two
independence scans over relevant SUSY parameter space by considering the constraints listed in Section II. These scans are
\begin{itemize}
\item Scan-III: Similar to Scan-I except that we fix $m_A = 800 {\rm GeV}$ and consider following parameter space
\begin{eqnarray}
 && 4 {\rm TeV} \leq m_{Q_2},m_{Q_3},m_{U_2}, m_{U_3}, m_{D_2}, m_{D_3} \leq 7 {\rm TeV}, \quad |A_t| \leq 6  \sqrt{m_{Q_3} m_{U_3}}, \nonumber \\
 &&   4 TeV \leq  m_{\tilde{g}}  \leq 10 TeV, \quad |\mu| \leq 20 TeV, \quad 1 \leq \tan\beta \leq 40,  \nonumber \\
 &&  -1 \leq \delta_{LL},\delta_{RR} \leq 1, \quad  -2.0\leq \delta_{LR},\delta_{RL} \leq 2.0, \quad -0.5 \leq \delta_{LR}^d,\delta_{RL}^d \leq 0.5.  \label{Scan-Region}
\end{eqnarray}
In our scan, we do not consider very large squark soft breaking parameters because in such a case,
the Higgs mass calculated by FeynHiggs suffers from large theoretical uncertainties.
\item Scan-IV: Similar to Scan-II except that the scan regions are now given by Eq.(\ref{Scan-Region}).
\end{itemize}

For both scans, we find the rate of $t \to c h$ may reach $10^{-8}$ in optimal case,
and a smaller value of $m_A$ can lead to a larger branching ratio. In Table \ref{table2},
we provide two benchmark points for future study with Point 5 obtained from Scan-III, and Point 6 from Scan-IV.
One can learn that, compared the prediction in heavy SUSY with that in low energy SUSY, although the optimal values of
$Br(t \to c h)$ is suppressed by at least two orders, they are still $10^6$ times larger than the corresponding SM prediction.

\section{Conclusion}

In this work, we studied the top quark FCNC decay $t \to ch$ in the MSSM
under the constraints from the Higgs mass measurement,
the LHC searches for sparticles, the vacuum stability, the precision electro-weak observables
and $B \to X_s \gamma$. From a  scan over the relevant parameter space, we found:
\begin{itemize}
  \item Due to the strong constraints from the measured Higgs mass and the results of SUSY searches
at the LHC, the branching ratio of $ t \to c h$ can only reach $O(10^{-6})$ in the MSSM, which is about
one order smaller than the old results.
  \item In the singlet extension of the MSSM, which can lift the Higgs mass at tree level,
$Br(t \to c h)$ can reach $O(10^{-5})$ in the allowed parameter space.
 \item The chiral-conserving mixings $\delta_{LL}$ and $\delta_{RR}$ can induce SUSY remanent effect on the rate of $t \to c h$.
For heavy squarks and gluino above $3 TeV$, $Br(t\to c h)$ can still reach $10^{-8}$.
 \end{itemize}

\vspace{0.2cm}

{\bf{Acknowledgement:}}

We thank Nima Arkani-Hamed, M.J. Herrero, Zhaoxia Heng, and Yosef Nir for helpful discussions.
This work was supported  by the National
Natural Science Foundation of China (NNSFC) under grant Nos. 10821504, 11222548, 11305049 and 11135003,
by Program for New
Century Excellent Talents in University, and also by the ARC Center of Excellence for Particle Physics
at the Tera-scale.

\end{document}